\documentclass{article}
\usepackage{spconf,amsmath,graphicx}
\usepackage{booktabs}
\usepackage[export]{adjustbox}
\usepackage{amsmath,graphicx,bm}
\usepackage{multirow}
\usepackage{subfigure}
\usepackage{cite}
\usepackage{graphicx}
\title{The 2020 Personalized Voice Trigger Challenge: Open Database, Evaluation Metrics and the Baseline Systems}
\name{Yan Jia$^{1}$, Xingming Wang$^{1,3}$, Xiaoyi Qin$^{1,3}$,Yinping Zhang$^{2}$,Xuyang Wang$^{2}$,Junjie Wang$^{2}$, Ming Li$^{1,3}$}
\address{$^1$Data Science Research Center, Duke Kunshan University, Kunshan, China \\
            $^2$AI Lab of Lenovo Research, Beijing, China \\
            $^3$School of Computer Science, Wuhan University, Wuhan, China }
\begin{document}
\ninept
\maketitle
\begin{abstract}

The 2020 Personalized Voice Trigger Challenge (PVTC2020) addresses two different research problems a unified setup: joint wake-up word detection with speaker verification on close-talking single microphone data and far-field multi-channel microphone array data. Specially, the second task poses an additional cross-channel matching challenge on top of the far-field condition. To simulate the real-life application scenario, the enrollment utterances are recorded from close-talking cell-phone only, while the test utterances are recorded from both the close-talking cell-phone and the far-field microphone arrays. This paper introduces our challenge setup and the released database as well as the evaluation metrics. In addition, we present a joint end-to-end neural network baseline system trained with the proposed database for speaker-dependent wake-up word detection. Results show that the cost calculated from the miss rate and the false alarm rate, can reach 0.37 in the close-talking single microphone task and 0.31 in the far-field microphone array task. The official website \footnote{https://www.pvtc2020.org/}and the open-source baseline system have been released. \footnote{https://github.com/lenovo-voice/THE-2020-PERSONALIZED-VOICE-TRIGGER-CHALLENGE-BASELINE-SYSTEM}

\end{abstract}
\section{Introduction}
\label{sec:intro}

Recently, speaker dependent voice trigger and wake-up word detection are gaining popularity among speech researchers and developers. It has been deployed in many real-life applications. With the contribution of deep learning, the performance of wake-up word detection and speaker recognition systems have improved remarkably in both close-talking and far-field scenarios.

The demand for authentication based on voice technologies, including keyword spotting (KWS) and text-dependent speaker verification (TDSV), is growing rapidly for personalized voice trigger devices. Generally, the KWS aims to detect a predefined keyword or a set of keywords in a continuous audio stream. Since the successful application of the Hidden Markov Model (HMM) in large vocabulary continuous speech recognition(LVCSR), the research of KWS system focuses on the statistical modeling. In recent years, various neural networks have been applied to KWS and achieved superior performance. HMM is utilized to construct the keyword model and the filler/background model, where the background model is trained with non-keyword speech, background noises and silence\cite{2,3,4}. The traditional Gaussian Mixture Models (GMM) were commonly used in statistic modeling for acoustic features in HMM-based approaches, and it is replaced by Deep Neural Networks (DNN) recently. End-to-end DNNs were applied to KWS and proved that DNNs perform well compared with HMM-based wake-up systems\cite{5}. Since then, more complex network structures have been adopted to build end-to-end KWS systems, including Convolutional Neural Networks\cite{Sainath2015ConvolutionalNN}, Recurrent Neural Networks\cite{10.1007/978-3-540-74695-9_23,WOLLMER2013252}, etc. On the other hand, with the success of deep learning in the speaker verification field\cite{8461375} and the demand for personalized trigger in smart home devices, the TDSV task has attracted much attention of speaker verification researchers.

However, the conventional wake-up word detection and speaker verification are carried out separately in the pipeline, where a wake-up word detection system is used to generate successful trigger followed by a speaker verification system used to perform identity authentication. In such a case, the wake-up word detection system and the speaker verification system are optimized separately, not through an overall joint optimization with a unified goal. As a consequence, their respective network parameters and extracted information are not effectively shared and jointly utilized. In recent studies, the combination of phoneme with speaker information has been the trend to improve the performance of the speaker verification system\cite{liu2018speaker,Wang2019}. The joint learning or multi-task learning network might be either very light at a small scale as a single always-on system, or with a much larger scale at the second stage after a successful wake-up by the first stage voice trigger. Therefore, there are still many open research fields that can be further explored for the joint modeling of wake-up word detection and speaker verification, including but not limited to:
\begin{itemize}
    \item Compact network design for personalized wake-up word detection with speaker verification
    \item Loss function design for personalized wake-up word detection with speaker verification
    \item Multi-task learning for personalized wake-up word detection with speaker verification
    \item Network structure design for personalized wake-up word detection with speaker verification
    \item Domain adaptation for personalized wake-up word detection with speaker verification
    \item Integration of wake-up word detection and speaker verification
    \item Speaker verification with segmentation from wake-up word detection
    \item Etc.
\end{itemize}

The 2020 Personalized Voice Trigger Challenge (PVTC2020), which aims at providing a common platform for the research community to advance the state-of-the-art, is a satellite event of the 12th International Symposium on Chinese Spoken Language Processing (ISCSLP 2021). The PVTC2020 challenge is focused on the speaker dependent wake-up word detection. We release a database named XIAO-LE\footnote{https://www.pvtc2020.org/dataDescription.html} containing recordings of wake-up words under the smart home scenario in this challenge. Besides, we also provide a two-stage speaker dependent KWS baseline system. When the KWS system triggers, we compare the trigger audio with the reference model created during the registration process and use another threshold to determine whether the sound that triggers the detector may be the wake-up word uttered by the registered user.

The paper is organized as follows. The details of XIAO-LE database are introduced in Section 2. Section 3 describes the challenge setup and the evaluation metrics. The adopted speaker dependent KWS baseline system is described in Section 4. Experimental results are shown in Section 5. Finally, the conclusion is provided in section 6.

\section{The XIAO-LE database}

The XIAO-LE database is provided by the AI Lab of Lenovo Research. It contains 658,995 utterances with 612 hours in total. The database covers 550 speakers and a wide range of channels from close-talking microphones to multiple far-field microphone arrays. It can be used for far-field wake-up word recognition, far-field speaker verification, and speech enhancement. 

The average duration of all utterances is around 3.8 seconds. During the recording process, recording devices, including two cell phones (16kHz, 16bit) and four microphone arrays (with 4 or 6 channels per array, 16kHz, 16bit), were set in a room designed as the real smart home environment.  

For the data collected by microphone arrays, each audio file has 4- or 6-channel signals, while for the data collected by cell phones, each utterance only has two channels signal. Recording devices and their corresponding distance information are shown in table \ref{tab:device_info}.

\begin{table}[htbp]   
	\caption{Distance of different recording devices.} 
	\label{tab:device_info}
	\center    
	\begin{tabular}{lcll}    
	\toprule   
	        Devices\_id  & Device and distance \\   
	\midrule   
			id1  & Cell phone, 0.2m \\
			id2  & Cell phone, 0.8m \\ 
			id3  & Microphone array, 1m \\
			id4  & Microphone array, 3m \\ 
			id5  & Microphone array, 3m \\
			id6  & Microphone array, 5m \\ 
	\bottomrule 
\end{tabular}  
\end{table}

\section{CHALLENGE DESCRIPTION}

\subsection{Task Setting}
Based on the XIAO-LE database, we have divided it into a training set, a development set, and two evaluation sets. Specifically, 300 speakers are selected for training, and  50 speakers are used as the development set. The rest speaker of the database is used for evaluation. The challenge provides two tracks for the participants, and the second task poses a cross-channel challenge: enrollment speech from the close-talk and test speech from far-field microphone arrays. In addition, to further simulate the real scenario, the text-independent content and confusing words also appear in trial files.

\subsubsection{Task 1: Joint wake-up word detection with speaker verification on close-talking data}
Only data collected by cell phones in the evaluation set from 100 speakers is adopted for performance evaluation in the first task. The evaluation set was separated into enrollment data and testing data. For each target speaker, the positive testing samples have `xiao le, xiao le' as a part of the speech, and it is indeed uttered from the target speaker. There might be some background noises or even other speakers’ voices before the target speaker says `xiao le, xiao le'. However, all utterances considered as positive samples have `xiao le, xiao le' uttered by the target speaker at the end of the speech in this case. Negative samples do not contain speech segments with the content `xiao le, xiao le' uttered by the target speaker. Note that utterances without `xiao le, xiao le' or with `xiao le, xiao le' that are not uttered by the target speaker are both considered as negative samples. 

\subsubsection{Task 2: Joint wake-up word detection with speaker verification on far-field multi-channel microphone array data}
For the second task, we adopt data from another 100 speakers in the evaluation set with no overlapping with the one in task 1. The evaluation data and the trials are constructed in the same way as task 1. The only difference is that the testing data only includes multi-channel synchronized audio data collected by one of those far-field microphone arrays.  Similar to task 1, `xiao le, xiao le' is always at the end of the sentence for all positive samples. In contrast, negative samples do not contain `xiao le, xiao le' or have speech segments `xiao le, xiao le' that are not uttered by the target speaker. The details about the trial design are shown in Table 2.

\begin{table*}[htbp]   
	\caption{Structure of the Trial files. Noting that, other text independent segments denote speech segments other than `xiao le, xiao le'} 
	\label{tab:trial}
	\center    
	\begin{tabular}{ccccc}    
	\toprule   
	        \multirow{3}*{\shortstack{Includes \\ `xiao le, xiao le'}}  &  \multirow{3}*{\shortstack{`xiao le, xiao le'part is from \\ the target speaker}} &  \multirow{3}*{\shortstack{Includes other text independent \\ segments from non-target \\ speakers before `xiao le, xiao le'}} &  \multirow{3}*{\shortstack{Includes other text \\
	        independent segments from 
	        \\ the target speaker}} & \multirow{3}*{Label} \\ 
	        & & & & \\
	        & & & & \\
	\midrule   
		 	yes  & yes & no & no & positive \\
			yes  & yes & no & yes & positive \\ 
			yes  & yes & yes & no & positive \\
			yes  & no & no & no & negative \\ 
			yes  & no & no & yes & negative \\
			yes  & no & yes & no & negative \\
			no  & n/a & n/a & yes & negative \\
			no  & n/a & n/a & no & negative \\
	\bottomrule 
\end{tabular}  
\end{table*}

\begin{table}[]
        \centering
        % \begin{tabular}{@{}lcccc@{}}
        \caption{Details about the Development and test set}
        \vspace{0.1cm}
        \begin{adjustbox}{max width=1 \columnwidth}
            \begin{tabular}{lccccc}
                \toprule
                  &               & utterances &  positive  & negative & enrollment\\ \midrule
                      & Task1    & 24.9k  & 3.6k  & 23.5k  & 1.6k\\
                Development  &  &        &   &  \\
                      & Task2   & 50.1k  & 4.6k  & 48.5k   & 1.6k\\ \midrule
                      & Task1         & 159.2k & 19.7k & 148.1k & 3.1k\\
                Evaluation        &  &        &   &  \\
                      & Task2 & 201.7k & 28.8k & 190.5k &3.1k\\ \bottomrule
            \end{tabular}
        \end{adjustbox}
        % \vspace{-0.6cm}
        \label{tab_dataset}
        \vspace*{-0.6cm}
\end{table}

\subsection{Design of trial files}

For speaker verification, participants can use up to three audios for enrollment for each speaker. The trial we provided contains three selected enrollment audios, one test audio, and the label which denotes whether the trial is positive or negative. In task 1, the utterances collected by cell phone in 0.2m are selected as the enrollment data, and those utterances collected by cell phone in both distances with 0.2m and 0.8m are used as the testing data. In task2, the utterances collected by cell phone in 0.2m distance are selected as the enrollment data, while utterances collected by multi-channel far-field microphone arrays are used for testing. We describe different scenarios in the trial construction with more details in Tab.\ref{tab:trial}. The correlation statistics are shown in Tab. \ref{tab_dataset}.

\subsection{Performance Measures}
In this challenge, we provide a leaderboard\footnote{https://www.pvtc2020.org/leaderboard.html} ranked by the metric $score_{wake\_up}$. The speaker dependent KWS performance of our baseline system, as well as systems submitted by participants in the challenge, is measured by matching statistics. The statistic score $score_{wake\_up}$ is calculated from the miss rate and the false alarm (FA) rate according to the following equation,
\begin{equation}
score_{wake\_up} = Miss + alpha * FA 
\end{equation}
Miss represents the proportion of errors in all positive label samples, and FA refers to the rate of errors in all negative label samples. The $alpha $ constant is set as 19, which is calculated by the assumption that the probability of the positive samples are 0.05. 

In addition, the real-time factor($F_{real-time}$)  is also evaluated as an auxiliary metric, which is calculated as the overall processing time of the evaluation trials on an Intel Core i5 core clocked at 2.6 GHz or similar processors divided by the total duration of all the testing samples. That is calculated as follows:
\begin{equation}
F_{real-time} = T_{process}(s) / T_{total\_test}(s) 
\end{equation}
$T_{process}(s)$ is the overall time cost of processing all the evaluation data in seconds, and $T_{total\_test}(s)$ is the total duration of the testing audios. In task2, multi-channel data will be considered as single-channel data when calculating $T_{total\_test}$. Besides, extracting the speaker embedding or features from the enrollment data is not counted in $T_{process}(s)$.
$F_{real-time}$ is a mandatory self-reported metric. Each submission is considered as a valid submission only when the corresponding self-reported real-time factor is lower than the given threshold. 

\section{The Baseline Systems}
\subsection{LSTM-based KWS system}
\label{sec:format}

\begin{figure}[th]
    \centering
    \includegraphics[width=9cm]{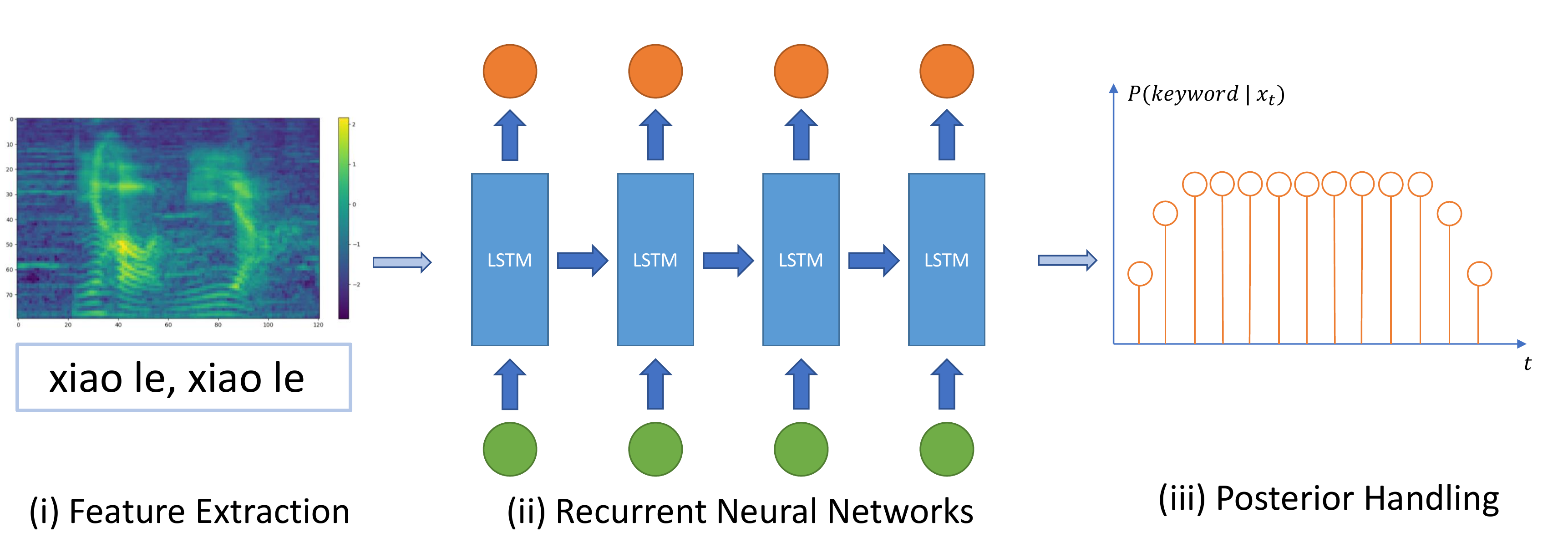}
    \caption{Framework of the baseline system.}
    \label{CNN_framework}
    % \vspace*{-0.6cm}
\end{figure}

This section presents our KWS baseline system, which is modified from the CNN-based KWS system \cite{Wu2020}. As shown in Figure \ref{CNN_framework}, our baseline system consists of three modules:(i) a feature extraction module, (ii) a stacked LSTM neural network and (iii) a confidence calculation module. 

The feature extraction module converts the audio signals into acoustic features. 80 dimensional log-mel filterbank features are extracted for speech frames with 25ms window size and 10ms window shift. Then we apply a segmental window with 40 frames to generate training samples that contain enough context information of sub-word as the input of the model. 
    
Our backbone network is constructed with a two-layer stacked LSTM structure, followed by an average pooling layer and a final linear projection layer.  For all LSTM layers, the hidden dimension is set to 128. A fully connected layer and a final softmax activation layer are applied as the back-end prediction module to obtain the subword occurrence probability of predefined keywords. Formally, suppose the input feature sequence of the model is $\bm{x} = (\bm{x}_1,\dots ,\bm{x}_t,\dots ,\bm{x}_T)$, and the output sequence of encoder is $\bm{h}=(\bm{h}_1,\dots,\bm{h}_t,\dots,\bm{h}_T)$, where $T$ equals 40,which is the length of the sequence. This encoder can be expressed as

\begin{equation}
    \bm{h} = LSTM(\bm{x})
\end{equation}

Then the average pooling layer generates a context vector $\bm{c}$:

% \begin{equation}
%     \alpha_{t} = \frac{1}{T}
% \end{equation}
\begin{equation}
    \bm{c} = \frac{1}{T} \sum_{t=1}^{T}\bm{h}_t
\end{equation}
Finally, we compute the sub-word, which are HMM states of keyword, occurrence probability of keyword by the fully connected layer and the softmax function:
\begin{equation}
    p_{w_i}(\bm{x}) = softmax(FC(\bm{c}))
\end{equation}
where $p_{w_i}(\bm{x})$ refers to the network output from input feature sequence $\bm{x}$ concerning sub-word $w_i$.

In the posterior handling module, while the acoustic feature sequence is projected to a posterior probability sequence of keywords by the neural network, we adopt the method proposed in \cite{liu2019loss, P2015Automatic} to make detection decisions. In this approach, we apply a sliding window with the length of $T_{conf}$ frames to compute detection scores and denote the input acoustic features in a window as $\bm{X}=\{\bm{x}_1, \bm{x}_2, \cdots \bm{x}_{T_{conf}}\}$. $\bm{w} = \{w_1, w_2 \cdots w_M\}$ represents the sub-words sequence of pre-defined keywords. Then the output confidence score $h(\bm{X})$ is computed by equation \ref{eq:confs},

\begin{equation}
    \label{eq:confs}
    h(\bm{X}) = \left [ \max \limits_{1 \leq t_1 < \cdots \leq T_{conf}} \prod_{i=1}^{M} p_{w_i}(\bm{x}_{ti}) \right ]^{\frac{1}{M}},
\end{equation}

where $p_{w_i}(\bm{x}_{ti})$ refers to the network output of $t^{th}$ frame at sub-word $w_i$. This method is suitable for the real-time situation. The system triggers whenever the confidence score is higher than the pre-defined threshold.

\subsection{Deep speaker embedding system}
\subsubsection{Data augmentation}
We adopt online data augmentation to improve the robustness of the speaker verification system and consider far-field microphone arrays data\cite{9036861}. We use the MUSAN \cite{snyder2015musan} and the RIRs-NOISES \cite{ko2017study} as the noise sources. The signal-to-noise-ratio(SNR) was set between 0 to 20 dB while pre-training and 0 to 15 dB while fine-tuning.

\subsubsection{Speaker verification network architecture}

The training process of the speaker verification baseline system is modified from the framework in \cite{chung2020in}. The whole architecture contains a front-end feature extractor, an encoding layer and a back-end classifier. We used ResNet34 \cite{he2016deep} with SE-block \cite{hu2018squeeze} as the feature extractor. The  attentive statistics pooling(ASP) \cite{okabe2018attentive} is adopt as the encoding layer. The ASP layer uses an attention mechanism to give different weights to different frames and generates a weighted average and a weighted standard deviation at the same time, which can effectively capture longer-term speaker feature variations. The AM-Softmax \cite{wang2018cosface} was set as the back-end classifier in the system. 

According to the experiments in \cite{qin2019far}, the strategy of transfer learning performs well in the far-field text-dependent speaker verification tasks. Therefore, we select the data from SLR38, SLR47\cite{primewords_201801}, SLR62, SLR82\cite{fan2019cnceleb}, SLR85\cite{himia} on openslr\footnote{http://openslr.org/resources.php} as the pre-training data. The specific data of the pre-training database are shown in Table \ref{sv_fine}. After that, we carry out fine-tuning on the XIAO-LE database. Fine-tuning schemes are divided into two types: the first is to use all the utterances of the pre-training database to construct a text-independent speaker verification system as a pre-training model; the second is to use only the database of XIAO-LE to fine-tune the pre-training model and get the target text-dependent system.

\begin{table}[h]
    \footnotesize
    \centering
    \caption{The specific data of database used for pre-training.}        
    \label{sv_fine}
    \begin{tabular}{lrrrr}
        \toprule
         & Speakers & Total hours & Utterances\\
        \midrule
        SLR38   & 855    & 100+ & 102600 \\
        \midrule
        SLR47   & 296    & 100+ & 50384 \\
        \midrule
        SLR62   & 600    & 200 & 237265 \\
        \midrule
        SLR82   & 274    & 1000 & 130108 \\
        \midrule
        SLR85   & 340    & 1500 & 108678 \\
        \bottomrule
    \end{tabular}
    \vspace*{0.5cm}
\end{table}

\subsection{Speaker dependent KWS system}

\begin{figure}[th]
    \centering
    \includegraphics[width=9.5cm]{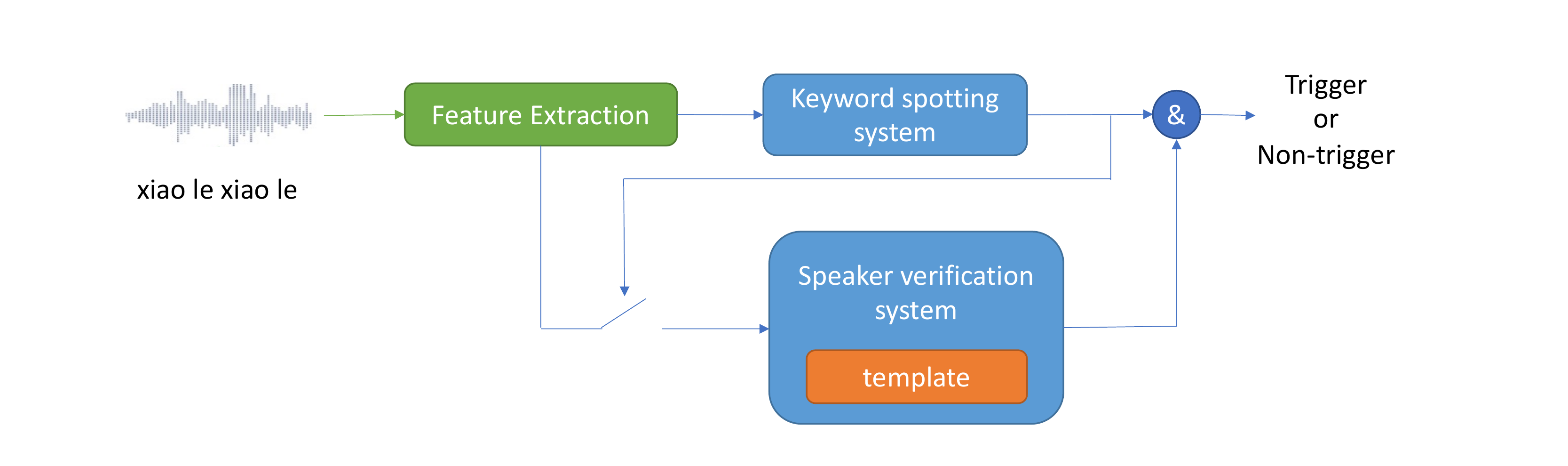}
    \caption{Framework of the baseline system.}
    \label{baseline_framework}
    % \vspace*{-0.6cm}
\end{figure}

Our baseline system consists of a wake-up system and a speaker verification system described above. As shown in Figure \ref{baseline_framework}, we designed a two-stage system that responds whenever the target speaker says the trigger phrase. When the KWS system triggers, the speaker verification system starts to decide whether the voice that triggers the detector is likely to be from the enrolled user. During enrollment stage, the average vector of the three utterances' embedding extracted from the target speaker is saved as the enrollment speaker embedding vector.

We compare any possible new `xiao le, xiao le' utterance with the stored templates as follows. The second stage detector produces timing information used to convert the acoustic feature sequence into a fixed-length vector. A separate, specially trained speaker verification network transforms this vector into a speaker embedding. We compare the cosine distances to the reference template created during enrollment with another threshold to decide whether the sound that triggers the wake-up word detector is likely to be the one from the enrolled user. This process prevents can help reduce the cases where the device is triggered by `xiao le, xiao le' spoken by another person and reduces the rate at which other, similar-sounding trigger phrases.

\section{Experimental results}

\subsection{Experiment setup}
\subsubsection{Keyword spotting system}
We determine target word labels by force-alignment with an LVCSR system. 
% Here, for keyword `xiao le, xiao le', we find out and center its on a window of 40 frames. 
For keyword 'xiao le, xiao le', the ending time of the first `xiao', the first `le', and the second `xiao' are found out and we center its on a window of 40 frames. 80 dims log fbank is adopted as our input acoustic features.  The KWS system is trained with cross-entropy loss. Stochastic gradient descent with Nesterov momentum is selected as the optimizer. The learning rate is first initialized as 0.01 and decreases by a factor of 0.1 whenever the model reaches a training loss plateau. We train the KWS model for 100 epochs with a batch size of 128 and employ early stopping when the training loss is not decreasing. In the evaluation period, we use a sliding window of 150 frames to compute the confidence score.

\subsubsection{Speaker verification system}
For pre-training, we also use Stochastic gradient descent as the optimizer. The initial learning rate is set as 0.01 and decreases by 0.1 per 20 epochs. The pre-trained model is trained for  50 epochs with a batch size of 256. For fine-tuning, the initial learning rate is set to 0.001 and the number of training epochs is set to 20. 

\subsection{Results}

All the experiments are evaluated on an Intel (R) Xeon (R) gold 5215 CPU clocked at 2.5 GHz.

\begin{table}[h]
    \footnotesize
    \centering
    \caption{Performances of the keyword spotting model on the dev set (the false rejection (FR) rate $[\%]$ under one false alarm (FA) per hour)}        
    \label{result_kws_baseline}
    \begin{tabular}{cccccc}
        \toprule
        Model & $F_{real-time}$ & Task1 & Task2 \\
        \midrule
        KWS baseline  & 0.09  & 2.00    & 5.11 \\
        \bottomrule
    \end{tabular}
    \vspace*{-0.4cm}
\end{table}

We choose the false rejection rate under one false alarm per hour as the KWS system performance criterion. Table \ref{result_kws_baseline} presents the KWS performance of the model regarding false rejection rate when the false alarm per hour is 1.

\begin{table}[h]
    \footnotesize
    \centering
    \caption{Performances of the speaker verification  model on the dev set (EER$[\%]$ and minDCF)}        
    \label{result_sv_baseline}
    \begin{tabular}{cccccccc}
        \toprule
        \multirow{2}*{Model}  & \multirow{2}{*}{$F_{real-time}$} & \multicolumn{2}{c}{Task1} & \multicolumn{2}{c}{Task2} \\
          \cmidrule(lr){3-4} \cmidrule(lr){5-6} & & EER & minDCF & EER & minDCF \\
        \midrule
        SV baseline & 0.11 & 1.32 & 0.16 & 1.90 & 0.21 \\
        % & EER  & 1.32    & 1.90 \\
        % SV baseline &     &  & \\   
        %           & minDCF & 0.16   &  0.21 \\
        \bottomrule
    \end{tabular}
    \vspace*{-0.4cm}
\end{table}

The results of our speaker verification system are shown in Table \ref{result_sv_baseline}. The threshold of the speaker verification system is determined by the development set. Two ways to determine the threshold have been used in our system. The first method is using the threshold of EER(Equal Error Rate) as the baseline version 1 system. The second is using the mean threshold of EER and minDCF\cite{greenberg20132012} denoted as $(threshold_{EER}+threshold_{minDCF})/2$, which greatly improves in the development set as baseline version 2 system. The results of the 
second method on development set are denoted as system V2 shown in Table \ref{result_baseline}

\begin{table}[h]
    \footnotesize
    \centering
    \caption{The $score_{wake\_up}$ of personalized KWS system on the test sets when alpha is 19}        
    \label{result_baseline}
    \begin{tabular}{ccccc}
        \toprule
        \multirow{2}*{Model}  & \multicolumn{2}{c}{Development} & \multicolumn{2}{c}{Evaluation} \\
        \cmidrule(lr){2-3} \cmidrule(lr){4-5} & Task1 & Task2 & Task1 & Task2 \\ 
        \midrule
        Personalized KWS baseline V1 & 0.19 & 0.33 & 0.75 & 0.78 \\
        Personalized KWS baseline V2 & 0.10 & 0.14 & 0.37 & 0.31 \\
        
        %         &  Task1 &  0.19  &  0.75  \\
        %   Personalized KWS baseline version 1 & & & \\
        %         & Task2 &   0.33 & 0.78   \\
        % \midrule
        %         &    Task1   &   0.10    &  0.37      \\
        % Personalized KWS baseline version 2  &  &   &  \\
        %         &   Task2    &   0.14    &   0.31     \\
        \bottomrule
    \end{tabular}
    \vspace*{0cm}
\end{table}

From Table \ref{result_baseline}, we can obtain the following observations from our baseline system. First, since the test recordings of task 2 are all far-field scene, the performance of the model on task 2 decreases significantly compared to task 1 on the development set. However, it is not the case in the evaluation set, which might be because multi-channel microphone array can compensate for the far-field condition. Second, the method to determine the threshold is an important factor affecting the final score.

\section{Conclusions}
In this paper, we introduce the setup of the ISCSLP 2020 Personalized Voice Trigger Challenge (PVTC2020) and describe the databases, tracks, rules, and baseline system of the challenge. The performance of the baseline system indicates that it is indeed a challenging task and we hope PVTC2020 can promote the advancement of research in the speaker dependent keyword spotting field. Our future works will focus on using the multi-task joint learning framework to handle the second stage detection.

\bibliographystyle{IEEEbib}
\bibliography{strings,refs}

\end{document}